\documentclass[prl,twocolumn,amsfonts,superscriptaddress,showpacs]{revtex4}
 
\usepackage[]{graphicx} 
\usepackage{amsbsy} 
\usepackage{float}
\usepackage{color}
\usepackage{subfigure}
\usepackage{amsmath}
 \usepackage{mathtools}

\newcommand{\be}{\begin{equation}}
\newcommand{\ee}{\end{equation}}
\newcommand{\bea}{\begin{eqnarray}}
\newcommand{\eea}{\end{eqnarray}}
\newcommand{\ba}{\begin{eqnarray*}}
\newcommand{\ea}{\end{eqnarray*}}

\begin{document} 
\title{Domain size effects on the dynamics of a charge density wave in 1T-TaS$_2$}

\author{G.~Lantz}
\affiliation{Institute for Quantum Electronics, Physics Department, ETH Zurich, CH-8093 Zurich, Switzerland}
\author{C.~Laulh\'{e}}
\affiliation{Synchrotron SOLEIL, L'Orme des Merisiers, Saint-Aubin, BP 48, 91192 Gif-sur-Yvette Cedex, France}
\affiliation{Univ. Paris-Sud, Universit\'{e} Paris-Saclay, 91405 Orsay, France}
\author{S.~Ravy}
\affiliation{Synchrotron SOLEIL, L'Orme des Merisiers, Saint-Aubin, BP 48, 91192 Gif-sur-Yvette Cedex, France}
\affiliation{Laboratoire de Physique des Solides, CNRS, Univ. Paris-Sud, Universit\'{e} Paris-Saclay, 91405 Orsay, France}
\author{M.~Kubli}
\affiliation{Institute for Quantum Electronics, Physics Department, ETH Zurich, CH-8093 Zurich, Switzerland}
\author{M.~Savoini}
\affiliation{Institute for Quantum Electronics, Physics Department, ETH Zurich, CH-8093 Zurich, Switzerland}
\author{K.~Tasca}
\affiliation{Institute for Quantum Electronics, Physics Department, ETH Zurich, CH-8093 Zurich, Switzerland}
\author{E.~Abreu}
\affiliation{Institute for Quantum Electronics, Physics Department, ETH Zurich, CH-8093 Zurich, Switzerland}
\author{V.~Esposito}
\affiliation{Swiss Light Source, Paul Scherrer Institut, CH-5232 Villigen PSI, Switzerland}
\author{M.~Porer}
\affiliation{Swiss Light Source, Paul Scherrer Institut, CH-5232 Villigen PSI, Switzerland}
\author{A.~Ciavardini}
\affiliation{Synchrotron SOLEIL, L'Orme des Merisiers, Saint-Aubin, BP 48, 91192 Gif-sur-Yvette Cedex, France}
\author{L.~Cario}
\affiliation{Institut des Mat\'{e}riaux Jean Rouxel - UMR 6502, Universit\'{e} de Nantes, 2 rue de la Houssini\`{e}re, F-44322 Nantes, France}
\author{J.~Rittmann}
\affiliation{Swiss Light Source, Paul Scherrer Institut, CH-5232 Villigen PSI, Switzerland}
\author{P.~Beaud}
\affiliation{Swiss Light Source, Paul Scherrer Institut, CH-5232 Villigen PSI, Switzerland}
\author{S.L.~Johnson}
\affiliation{Institute for Quantum Electronics, Physics Department, ETH Zurich, CH-8093 Zurich, Switzerland}
\date{\today} 
 
\begin{abstract} 
Recent experiments have shown that the high temperature incommensurate (I) charge density wave (CDW) phase of 1T-TaS$_2$ can be photoinduced from the lower temperature, nearly commensurate (NC) CDW state. Here we report a time-resolved x-ray diffraction study of the growth process of the photoinduced I-CDW domains. The layered nature of the material results in a marked anisotropy in the size of the photoinduced domains of the I-phase. These are found to grow self-similarly, their shape remaining unchanged throughout the growth process. The photoinduced dynamics of the newly formed I-CDW phase was probed at various stages of the growth process using a double pump scheme, where a first pump creates I-CDW domains and a second pump excites the newly formed I-CDW state. We observe larger magnitudes of the coherently excited I-CDW amplitude mode in smaller domains, which suggests that the incommensurate lattice distortion is less stable for smaller domain sizes.
\end{abstract}

\pacs{XXXXXXX; } 
\maketitle

\section{Introduction}

1T-TaS$_2$ is  a very well studied and prominent example of a strongly correlated material, which exhibits a rich phase diagram that contains three different charge-density-wave (CDW) phases, a Mott insulator phase and even a superconducting phase\cite{Sipos2008a}. Excitation of 1T-TaS$_2$ with intense femtosecond laser pulses has been shown to stimulate transitions between these states \cite{Perfetti2006a,Eichberger2010,Hellmann2010a,Sun2015} and, in some circumstances, can also lead to formation of a new ``hidden" metastable phase with properties distinct from those accessible at thermodynamic equilibrium\cite{Stojchevska2014}. Understanding the dynamics of these photoinduced transitions and how they relate to their equilibrium thermodynamic counterparts is of key interest in understanding the competing mechanisms of electronic correlations and electron-phonon interaction that result in so many different kinds of stable and metastable orderings in this material. 

\begin{figure}
\includegraphics[angle=0,width=1\linewidth,clip=true]{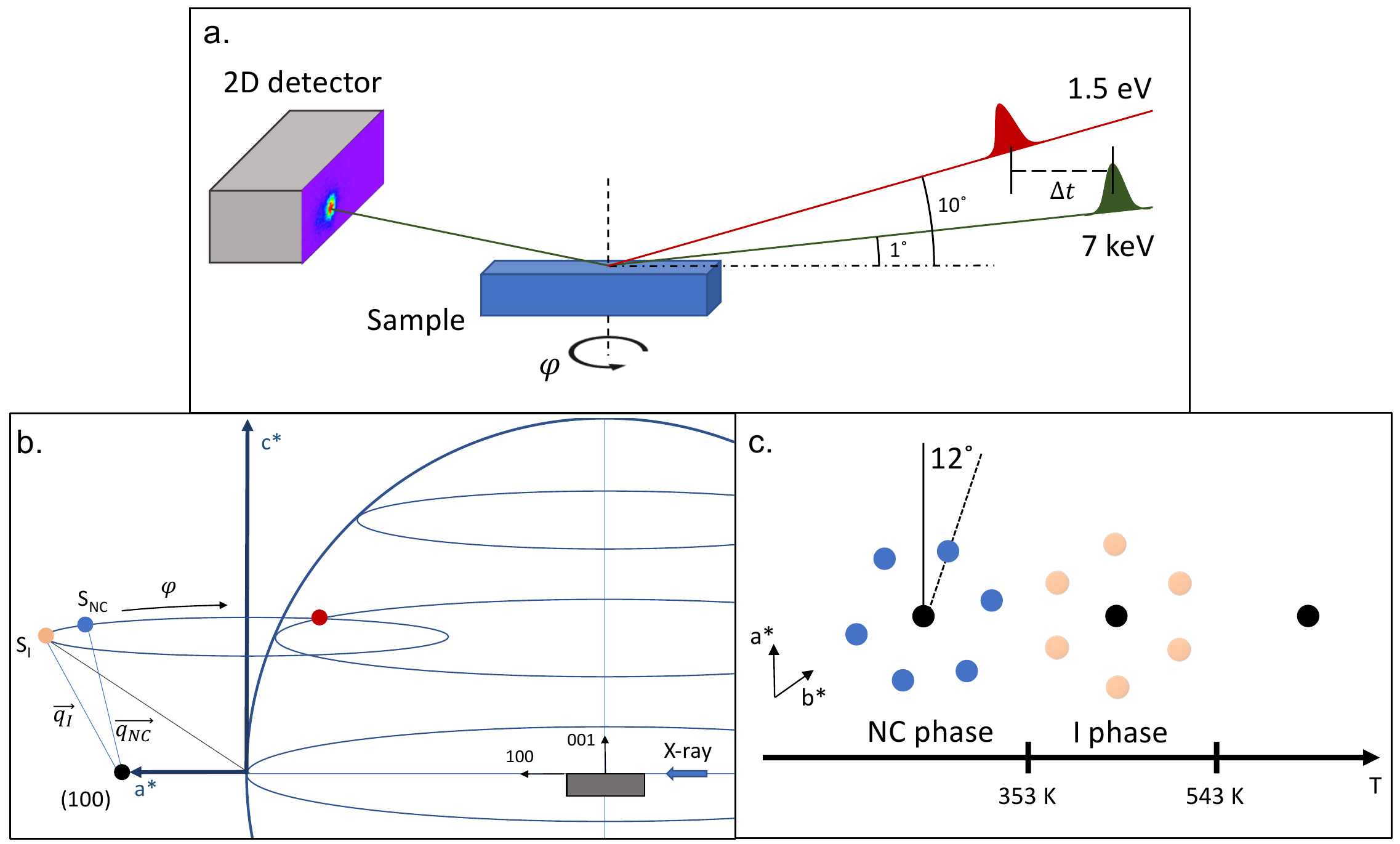}
\caption{Experimental setup. \textbf{a.} Optical pump x-ray probe experimental setup. \textbf{b.} Ewald sphere for the NC and I satellite peaks. The rotation scan around the surface normal ($\varphi$) is represented. \textbf{c.} Schematic representation of the superlattice peaks in the ab plane as a function of temperature. } 
\label{ewald} 
\end{figure}

1T-TaS$_2$ has a layered structure composed of S-Ta-S sheets, where the Ta atom is octahedrally coordinated by six S atoms. At high temperature the material is metallic with no CDW\cite{Sipos2008a}. Below 543~K, an incommensurate (I) CDW arises, characterized by modulation vectors $\vec{q_I}=0.283\vec{a^*}+\frac{1}{3} \vec{c^*}$ and equivalents due to the 3-fold symmetry along the c axis. Below 353~K there is a phase transition to the nearly commensurate (NC) phase, in which the modulation wave vectors change to $\vec{q_{NC}}=0.245\vec{a^*}+0.068\vec{b^*}+\frac{1}{3} \vec{c^*}$ and equivalents due to the 3-fold symmetry \cite{Spijkerman1997}. The main difference between the I and NC phases is a rotation of 12$^\circ$ of the CDW within the ab plane as represented in Fig. \ref{ewald} \textbf{c}. The NC phase has a regular domain-like structure where small commensurate domains with wave vector $\frac{3}{13} \vec{a^*}+\frac{1}{13}\vec{b^*}$ are separated from each other by a discommensuration regions. Below 180~K, these discommensuration regions disappear and the material becomes insulating, forming the commensurate (C) phase. Femtosecond laser excitation has been shown to trigger transitions from the C and NC phases to the NC and I phases, respectively~\cite{Han2015,Laulhe2015}. The photoinduced NC to I transition is the focus of the present work.

Previous experiments have shown that the correlation length of the photoinduced I phase increases with time, up to several nanoseconds after the transition is triggered by a pump pulse \cite{Laulhe,Vogelgesang2017}. According to our current understanding of this process, initially several independent regions exhibiting I-CDW phase modulations nucleate and grow from the excited NC phase. These regions continue to grow until they coalesce, forming a mutlidomain I-CDW phase. The domain size then slowly increases through a coarsening process, following the universal Lifshitz-Allen-Cahn growth law (t$^{1/2}$) \cite{Laulhe}. The above mentioned studies focused on the growth of CDW correlation length exclusively in the ab plane directions. Adding a description of domain growth in the c directions is the first aim of the present work. Another question raised by the previous experiments is the nature of the I-CDW when domain sizes are small, before the coarsening process has taken place. Induced static defects have been used to study the domain size dependence of CDW states, with results suggesting that smaller domains lead to larger CDW amplitudes and a discretization of the nesting vector \cite{Zhang1996,Kim2001}. 
In these studies it is not clear, however, whether the observed changes of the CDW properties originate from the induced defects or from the domain sizes. The photoinduced NC-to-I phase transition gives an unique opportunity to study the properties of a CDW for different domain sizes while the defect density remains the same. The second aim of the present work is to obtain information on the amplitude mode of the photoinduced I-CDW, by photoexciting and probing its dynamics.

Although optical reflectivity has been used in the past to track the dynamics of the CDW in the various phases of 1T-TaS$_2$~\cite{Demsar2002a}, it cannot to clearly distinguish the dynamics of the residual NC phase from those of the slowly forming I phase. Due to its ability to resolve different regions in reciprocal space, time-resolved x-ray diffraction can, on the other hand, distinguish them clearly and independently track their dynamics. Moreover, the high energy resolution available at synchrotron beamlines allows a precise reciprocal space mapping of the satellite peak arising from the periodic lattice displacement (PLD). Its intensity, the position, and shape can then be monitored as a function of delay after laser excitation. While the intensity provides information on the ratio between the I and NC phases, the position relates to the modulation vector. The satellite peak width can be determined in various directions and used to derive the correlation lengths along those directions. Previous experiments using electron diffraction or point detectors have been able to study the intensity and the correlation length of the CDW only in the in-plane directions \cite{Vogelgesang2017,Haupt2016,Laulhe}.

In this article we study 1T-TaS$_2$ using time-resolved x-ray diffraction to observe the dynamics of the NC-to-I photoinduced phase transition. We first focus on the growth of the correlation length of the domains along the different crystal directions using a 3D reciprocal space mapping. The time evolution of the intensity and the correlation lengths are compared to the previous experiments. Secondly we use a double-pump scheme to study the properties of the incommensurate CDW for various domain sizes.

\section{Experimental Method}
The 1T-TaS$_2$ samples are platelet-like, (001)-oriented single crystals of 5 mm diameter and a few hundred microns in thickness. They were synthesized using an iodine transport method as reported in Ref. \onlinecite{Ravy2012}. The sample temperature is controlled using a nitrogen cryoblower.

\begin{figure}
\includegraphics[angle=0,width=1\linewidth,clip=true]{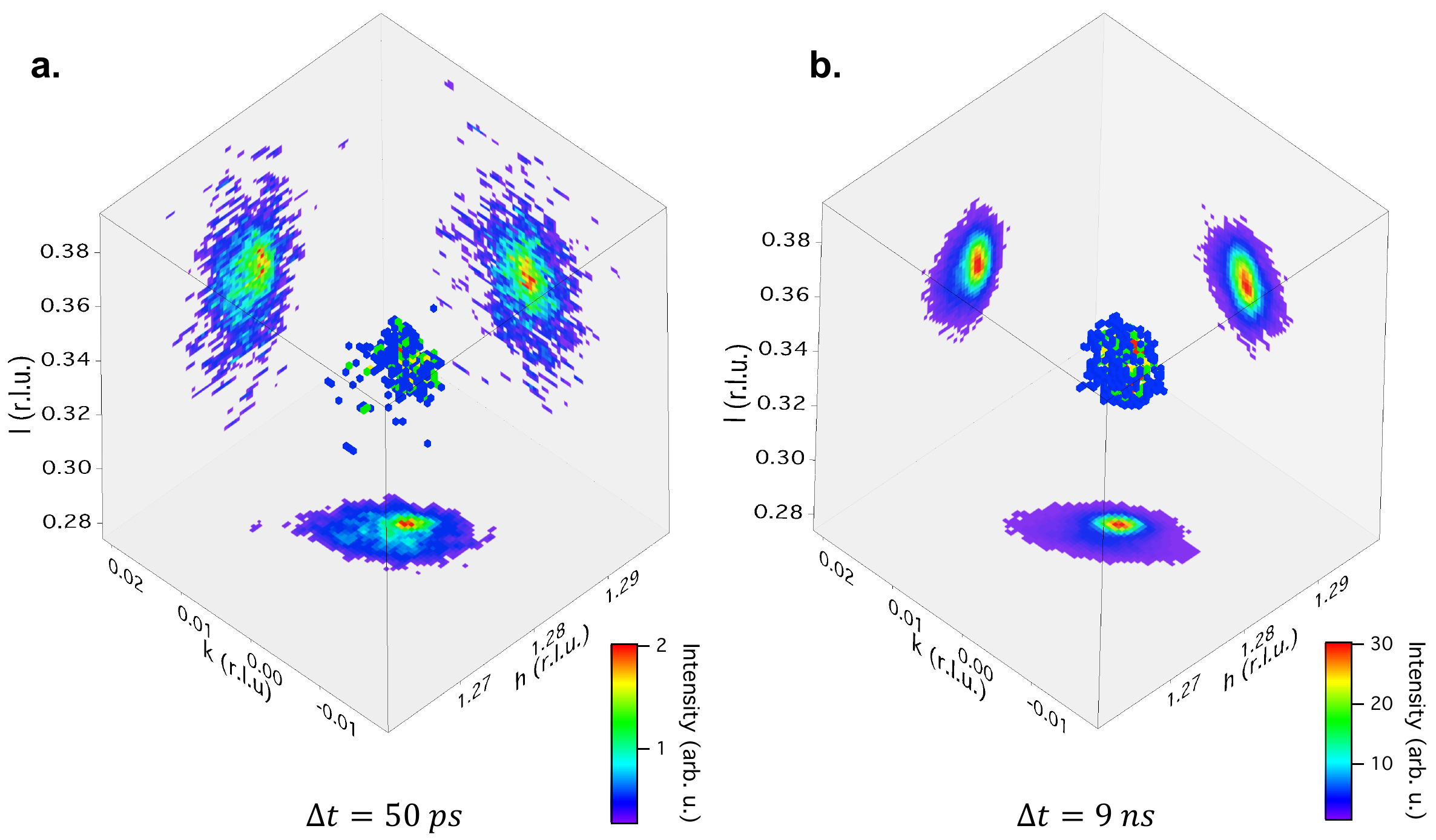}
\caption{3D representation of the photoinduced I-phase superlattice peak in reciprocal space. The projections in all three planes are represented for \textbf{a.} 50 ps delay and \textbf{b.} 9 ns delay. The color scale is adapted for the two delays but the relative numbers are consistent.} 
\label{3d} 
\end{figure}

For the mapping of reciprocal space during the growth of the photoinduced I phase, we perform time-resolved x-ray diffraction after excitation from a near-infrared pump on the microXAS beamline at the Swiss Light Source (PSI, Villigen). The pump is a 100~fs long 1.55~eV pulse from a regenerative Ti:Al$_2$O$_3$ amplifier operating at 1~kHz repetition rate that is synchronized to the electron bunches of the storage ring~\cite{Gawelda2007}.  The x-ray probe pulses arise from the isolated electron bunch of the hybrid mode filling pattern, and exhibit a duration of 70 ps \cite{Milas2010}. A photon energy of 7 keV is selected by means of a Si(111) double monochromator, leading to a high momentum resolution of 3.5 10$^{-4}$ \AA$^{-1}$. The latter value is two orders of magnitude finer than with time-resolved electron diffraction \cite{Vogelgesang2017,Haupt2016}. This configuration was chosen for the reciprocal space mapping since the dynamics of domain growth occur on timescales of hundreds of picoseconds, and a very good momentum resolution and high flux are necessary. A grazing incidence geometry is used with an angle with respect to the sample's surface of 1$^\circ$ for the x-ray probe and 10$^\circ$ for the optical pump. The attenuation depths in intensity of the pump and the probe are 30 nm and 130 nm, respectively \cite{Beal1975}. The scheme of the experimental setup is represented in Fig. \ref{ewald} \textbf{a.} We use a 2D PILATUS detector operated in a gated detection mode and perform scans as a function of rotation angle $\varphi$ about the surface normal, which keeps the attenuation depth of the probe and the pump constant. $\varphi$ scans make the diffracted peak traverse the Ewald sphere and enable a 3D reconstruction of the peak, as shown in Fig. \ref{ewald} \textbf{b.}. The sample was pre-oriented on the CRISTAL Beamline at SOLEIL synchrotron (Saint-Aubin, France), where an orientation matrix was computed. The orientation matrix was confirmed during the time-resolved experiment using the (111) and the $\vec{a^*}+\vec{q_{NC}}$ reflections. An image is recorded for every $\varphi$, which by using the orientation matrix and angles enables us to attribute a hkl value to each pixel. Therefore the reciprocal space mapping of the satellite peak can be analyzed in 3D, as represented in Fig. \ref{3d}. 

\begin{figure}
\includegraphics[angle=0,width=0.8\linewidth,clip=true]{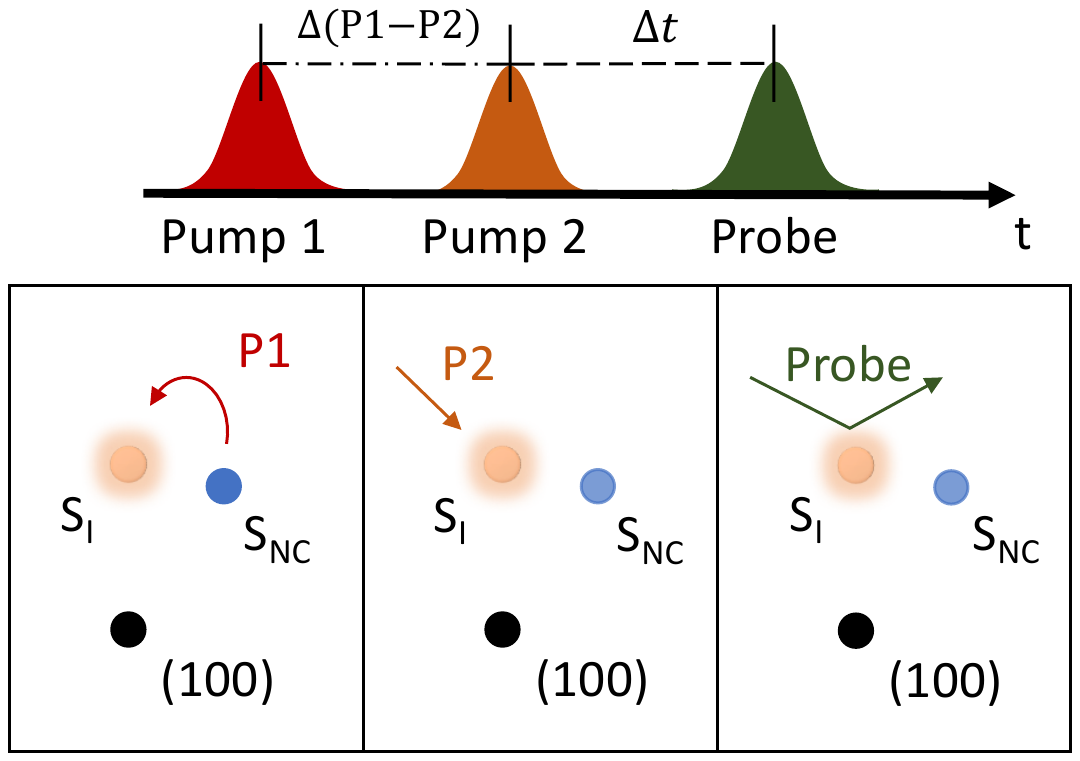}
\caption{Double pump scheme. The first pump creates the I-phase whereas the second pump perturbs the I-phase in order to probe its dynamics.} 
\label{dpp} 
\end{figure}  

For studying the femtosecond dynamics of the newly formed I phase during the domain growth process, a better temporal resolution is required.  For this portion of our experiments we use the electron-beam slicing operation mode, which produces x-ray pulses of 120 fs duration from the storage ring at a repetition rate of 2~kHz~\cite{Beaud2007}. Relative to the picosecond duration unmodified x-ray beam from the synchrotron, the flux of this sliced beam is reduced significantly.  To mitigate this loss, we sacrifice some energy resolution by replacing the Si(111) monochromator with a multilayer monochromator which increases the energy bandwidth to about 1.3\%. The diffracted x-rays from the sample are measured with an avalanche photodiode point detector. The pump scheme used to investigate the I-phase dynamics is shown in Fig. \ref{dpp}. We employ two pump pulses P1 and P2 separated by a delay $\Delta(P1-P2)$.  The P1 pulse triggers formation of the I phase, while the P2 pulse re-excites the sample to stimulate dynamics within the I phase induced by P1.  
We then probe the amplitude of the incommensurate modulation by monitoring the diffracted intensity from the I satellite peak as a function of delay time $\Delta t$ relative to the P2 pulse.
Since the growth of the domains takes place on a timescale of hundreds of picoseconds, it can be assumed that the relative volume of the photoinduced I phase is constant in the $\Delta t$ range covered [0 - 5 ps]. In this case, any variation of the intensity of the I satellite peaks reflects a change of the amplitude of the PLD of the I-phase. By varying $\Delta (P1-P2)$ we modify the size of the I-phase domains. Since the growth of the domains is on a timescale of hundreds of picoseconds, the intensity of the I-phase can be considered constant over the 0-5 ps timescale. In order to get a precise measure of $\Delta (P1-P2)$, we measure x-ray diffraction as a function of delay between the pump beams and the x-rays from the NC peak with the P1 and P2 pumps alternately blocked. Since the overlap of the x-rays and pump coincides with a sudden decrease of the NC peak intensity~\cite{Laulhe}, this gives a precise measure of the delay between P1 and P2.

All measurements were performed for the two temperature setpoints 230 K and 265 K. However, it is important to note that at a repetition rate of 1 kHz, the laser-induced average heating of the sample can bring the base temperature close to the critical temperature of the NC-I phase transition. For each of the temperature setpoints, we paid particular attention to decrease the laser fluence until no intensity at the I satellite peak position was present for negative pump-probe delays. The resulting absorbed fluences are 10.5 mJ cm$^{-2}$ at 230 K, and 4.6 mJ.cm$^{-2}$ at 265 K. In these conditions, one can consider that the actual temperature lies just below the NC-I transition temperature, 353 K. This allows us to determine the laser-induced temperature increase from the set temperature: 123 K for measurements taken at a set temperature of 230 K and at high fluence, and 88 K for measurements taken at a set temperature of 265 K and at low fluence. The same difficulty was encountered also in the work reported in Ref \citep{Laulhe}.

\begin{figure}
\includegraphics[angle=0,width=1\linewidth,clip=true]{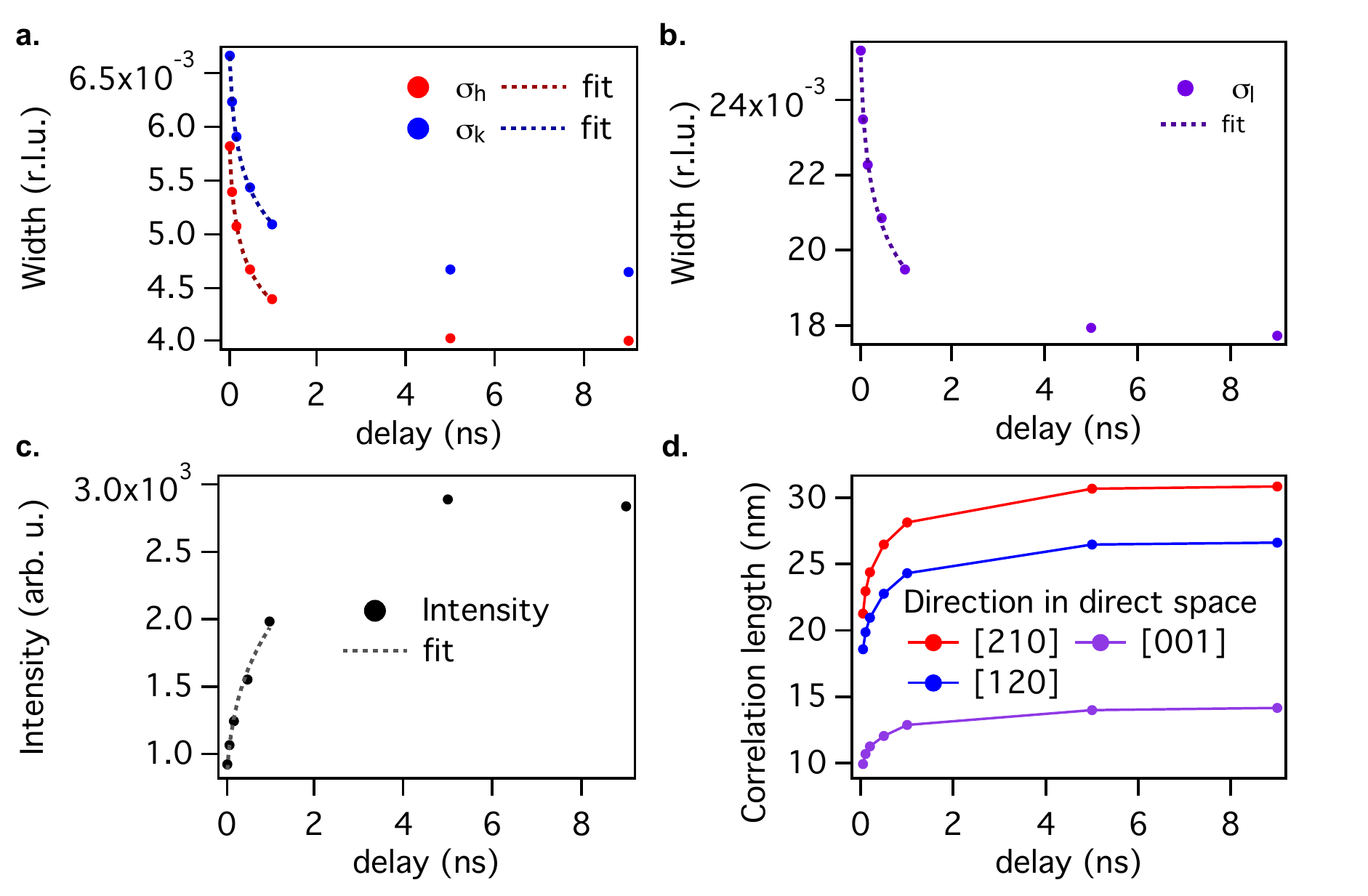}
\caption{Intensity and width of the I-satellite peak versus delay obtained at 230~K for an absorbed fluence of 10.5~mJ~cm$^{-2}$. \textbf{a.} Widths in the hk plane. \textbf{b.} Width in the l direction \textbf{c.} Integrated intensity. The widths and intensity are fitted with a power law in the 0-1~ns delay range. \textbf{d} Correlation length of the I-CDW versus time delay along different real space directions. The widths measured in the different directions exhibit the same relative changes as a function of $\Delta t$, but different absolute values.} 
\label{230K_90mW} 
\end{figure}   

\begin{figure}
\includegraphics[angle=0,width=1\linewidth,clip=true]{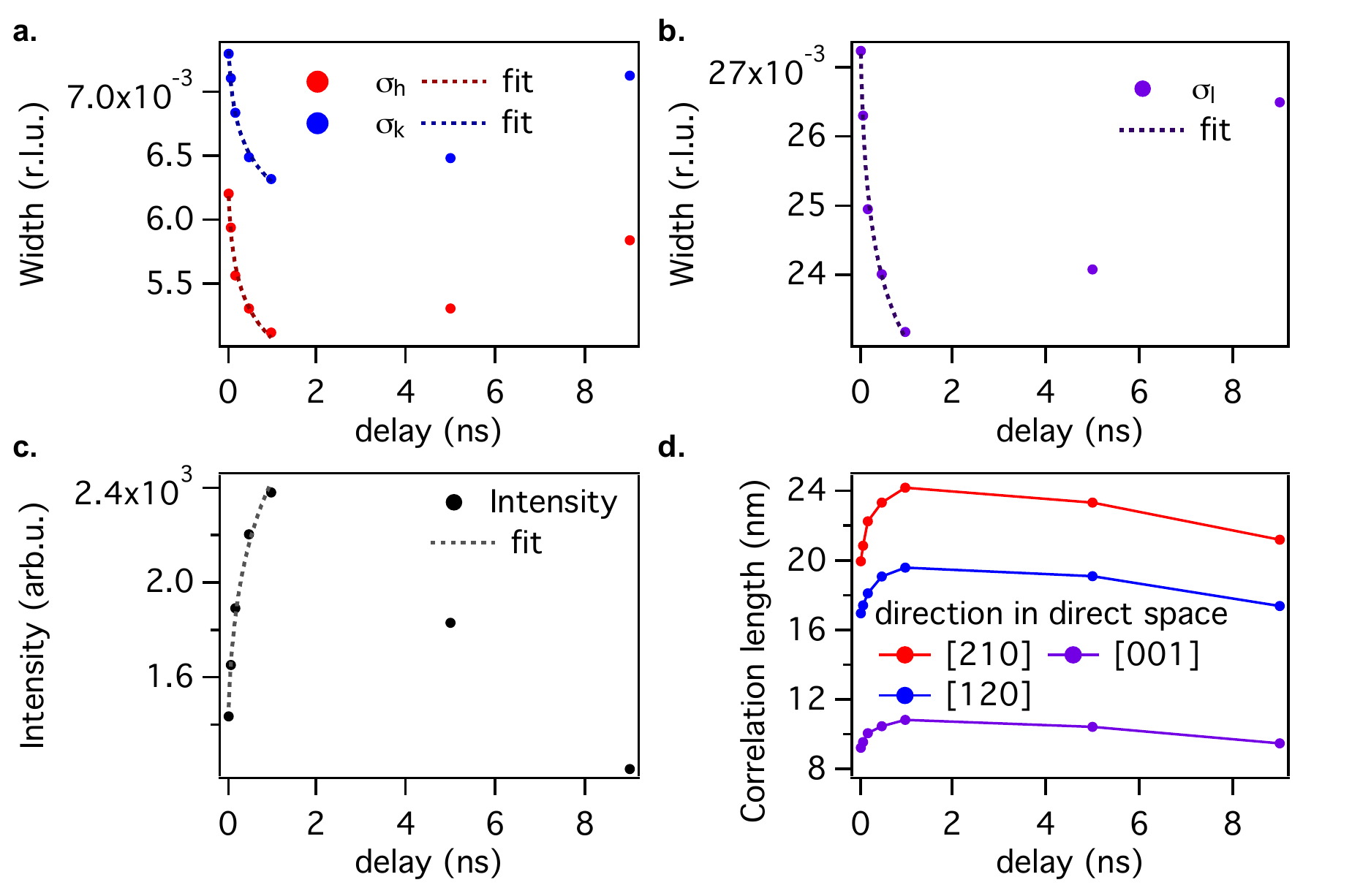}
\caption{Intensity and width of the I-satellite peak versus delay obtained at 265~K for  a fluence of 4.6~mJ~cm$^{-2}$. \textbf{a.} Widths in the hk plane, \textbf{b.} Width in the l direction \textbf{c.} Integrated intensity. The widths and intensity are fitted with a power law in the 0-1~ns delay range. \textbf{d} correlation length of the I-CDW versus delay along different real space directions.} 
\label{265K_40mW} 
\end{figure}  

\begin{figure}
\includegraphics[angle=0,width=1\linewidth,clip=true]{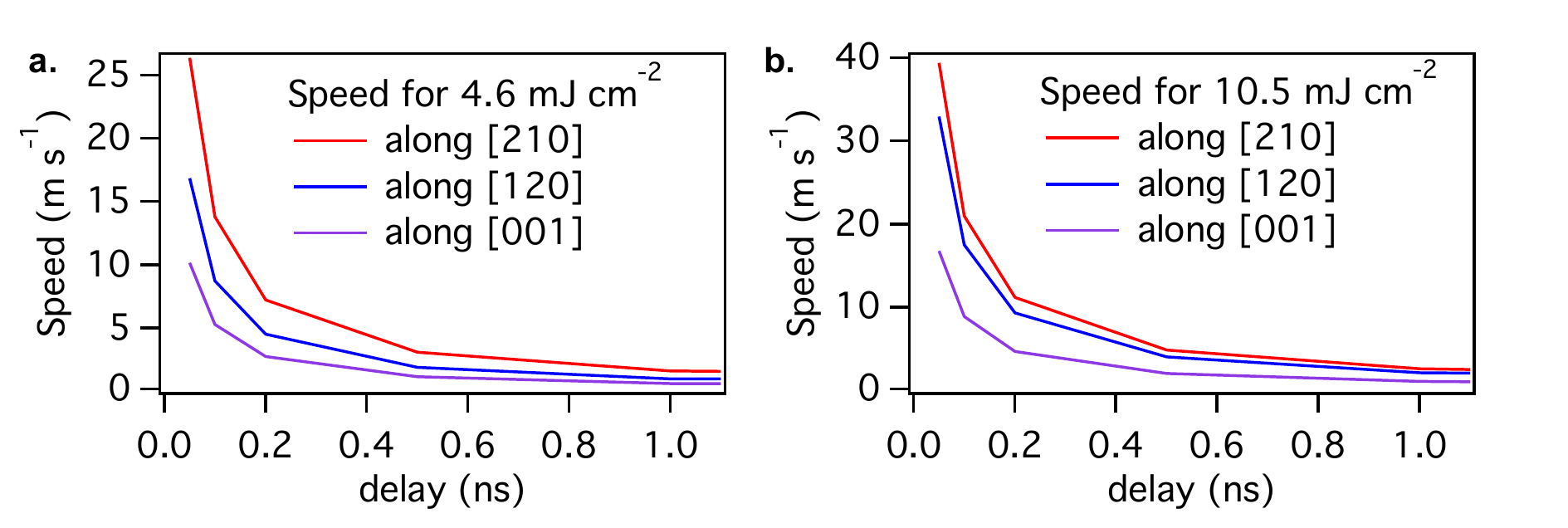}
\caption{Growth velocity  along different real space directions versus delay \textbf{a.} for 10.5~mJ~cm$^{-2}$ and \textbf{b.} for 4.6~mJ~cm$^{-2}$} 
\label{speed} 
\end{figure}  

\section{Results}

We measured the diffracted intensity around the I satellite peak position (1.283  0  0.333), for various delays after laser excitation, starting from the NC phase at 230 K.  A reciprocal space mapping was carried out for each of the 7 pump-probe delays investigated: 50~ps, 100~ps, 200~ps, 500~ps, 1~ns, 5~ns, and 9~ns. Fig. \ref{3d} shows the 3D representation of the I-satellite peak for 50~ps and 9~ns with the projections on the hk, kl, and hl planes. The color scale is adapted to the intensity of the peak, which is ten times lower for the 50~ps delay. The peak at 9~ns is found to be much narrower than at 50~ps, in agreement with previous experiments~\cite{Haupt2016,Laulhe}.

In order to better quantify the shape of the peak using a parameter-free model, the first three moments of the distribution are calculated: the zeroth moment is the integrated intensity, the first moments give the peak position, and the second moments give the variances (widths) along the h, k and l directions. The first moment does not change significantly with delay within our resolution. The variances $\sigma_h$, $\sigma_k$, and $\sigma_l$ of the distribution of diffracted intensity are shown in panels a and b of  Fig. \ref{230K_90mW}a and \ref{230K_90mW}b as a function of delay time, the integrated intensity is given in panel c. We observe that the integrated intensity increases up to 5 ns and then stabilizes, while the widths in all three directions decrease and stabilize on a similar timescale. 
As discussed in the appendix, the instrument resolution makes a negligible contribution to the observed widths and therefore can be neglected.  
 
To compare these results to previous experiments, it is helpful to convert the peak widths sigma to correlation lengths of the forming I-CDW state. If we suppose that the domain dimensions follow a Gaussian distribution, the 
correlation lengths are given by $\xi_{a^*} =2\pi/(2ln(2) \sigma_h a^*)$, $\xi_{b^*} =2\pi/(2ln(2) \sigma_k b^*)$, and $\xi_{c^*} =2\pi/(2ln(2) \sigma_l c^*)$ in the $a^*$, $b^*$ and $c^*$ directions of the reciprocal space, which corresponds to the [210], [120] and [001] crystallographic directions. The correlation length along $c^*$ is twice smaller than the ones measured in the ($a^*$, $b^*$)plane, which is reminiscent of the layered nature of the material. Their dynamical behaviors, however, are very similar: for example, the relative change in width between 1 ns and 50 ns after excitation $(\sigma(50 ps)-\sigma(1 ns))/\sigma(50 ps)\approx0.23(1)$ is approximately the same along all directions.

In order to confirm that this analysis does not depend on the basis in reciprocal space selected for the analysis, the in-plane widths are also analyzed in a cylindrical basis with the center on the (100) Bragg reflection, the cylindrical axis along the c$^*$ axis, and the polar axis along a$^*$. This particular basis is chosen because the phase transition from NC to I phase corresponds  to an in-plane rotation of the wavevector. The ratios and delay trend using these cylindrical coordinates are identical for both the radius and azimuthal angle, which is consistent with the explanation of Refs. \citep{Laulhe,Haupt2016} where the authors argue that the appearance of the I-phase arises from a highly disordered phase and not from a pure rotation.

In the experimental conditions used in the work of Ref.~\cite{Laulhe} , the integrated intensity of the I satellite peak is constant in the 100 - 500 ps pump-probe delay range, which is a signature of coarsening phenomenon. In the present work ,  the I satellite peak intensity keeps growing continuously with time ( Fig.~ \ref{230K_90mW}c), which indicates that the photoinduced I-phase regions grow independently, without contact. We apply a power law fit (A$\Delta t^\alpha$) over the 0-1 ns delay range in order to quantitatively describe the time-dependent behavior of both the intensity and widths. Results are shown as dashed lines in Fig.~\ref{230K_90mW}a-\ref{230K_90mW}c, and the fitting parameters are listed in Table \ref{tconst}. For an excitation fluence of 10.5 mJ/cm$^2$, the three widths fit to a power law with $\alpha \approx -0.09$. The volume of the domains is calculated as proportional to the product of the correlation lengths, which is also a power law with an exponent equal to the sum of the three exponents. The sum of the exponents is 0.26~$\pm$~0.01 which is the same exponent as the power law describing the intensity growth, within error bars. Therefore the intensity is proportional to the volume of the domains, which further show that with the present experiment we do not observe domain coarsening but rather the growth of the photoinduced I-phase regions.

The same $\varphi$ scans were performed at 265 K with a lower absorbed fluence of 4.6~mJ~cm$^{-2}$. The widths and the intensity are analyzed using the same procedure as for the lower temperature, see Fig. \ref{265K_40mW}. We obtain quite similar results as at 230 K. The dynamics of the widths and the ratio $(\sigma(50 ps)-\sigma(1 ns))/\sigma(50 ps)\approx0.14(1)$ are the same for all three directions., and the sum of the different exponent is 0.171~$\pm$~0.01, which corresponds to the exponent of the intensity. The intensity decreases after 1~ns, which does not impact the fits since the range used was 0-1~ns (see Table \ref{tconst}). This drop is much more pronounced than the increase of the widths, suggesting that the regrowth of the NC phase may first act to suppress I phase domains with smaller dimensions.

\begin{table}
\begin{tabular}{|p{1.3cm}|c|c|c|c|c|}
  \hline
  Fluence mJ~cm$^{-2}$& Int & $\sigma_h$ & $\sigma_k$ & $\sigma_l$ &Vol\\
  \hline
 4.6 & 0.166(10) & -0.066(5) & -0.050(2) & -0.055(3)&-0.17(1)\\
 10.5 & 0.261(17) & -0.093(3) & -0.089(2) & -0.085(4)&-0.26(1)\\  
  \hline
\end{tabular}
\caption{Exponent of the power law fits, A$\Delta t^\alpha$,  of the intensity and the three widths, for two fluences. The exponent that describes the time evolution of the I phase volume is calculated by summing the three width exponents.}
\label{tconst} 
\end{table}

\begin{figure}
\includegraphics[angle=0,width=1\linewidth,clip=true]{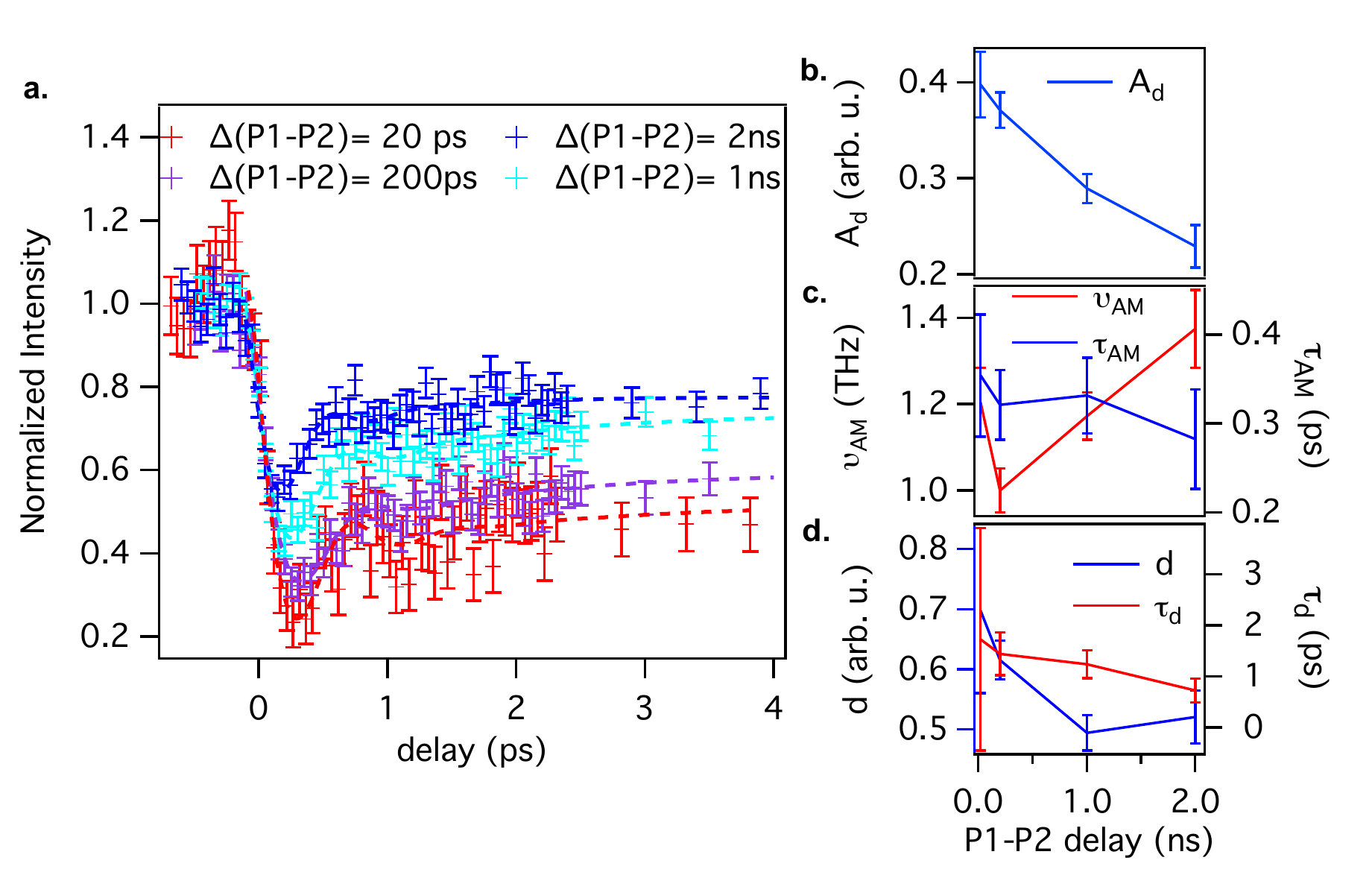}
\caption{Double pump experiment results. \textbf{a.} Time evolution of the intensity of the I satellite peak versus delay between P2 and probe ($\Delta t$), for various P1-P2 delays. \textbf{b.-c.} Refined values of the fit parameters within the adapted displacive excitation model (see text), for different P1-P2 delays. The amplitude of the displacement is very sensitive to the P1-P2 delay as well as the final displacement.} 
\label{expdpp} 
\end{figure}  

\begin{figure}
\includegraphics[angle=0,width=0.8\linewidth,clip=true]{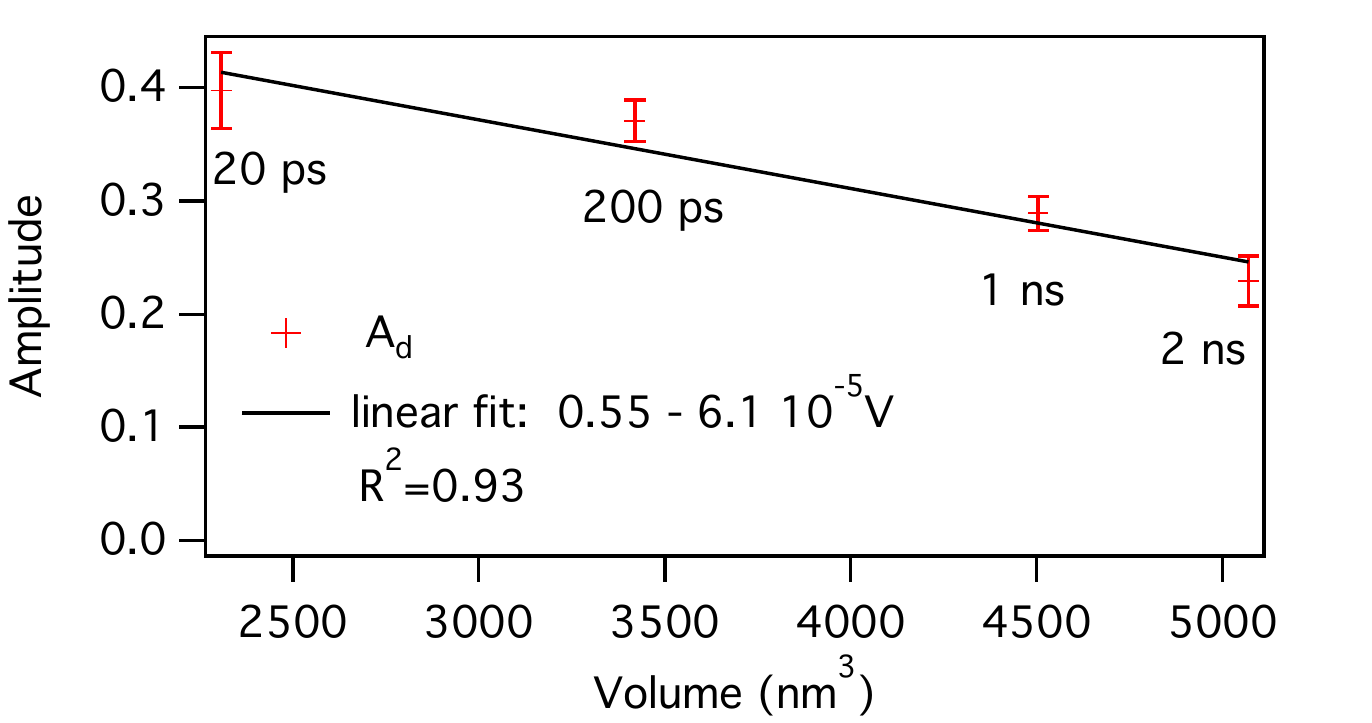}
\caption{Magnitude of the coherently excited amplitude mode versus domain volume of the I phase for the double pump experiment. The relation between the magnitude and the volume is linear.} 
\label{widthdpp} 
\end{figure}

We now turn to the dynamics of the I-phase for different domain sizes. Using the double pump setup, we scanned $\Delta t$ for various $\Delta (P1-P2)$, at the position in $\varphi$ that maximize the intensity of the emergent I phase peak for various $\Delta (P1-P2)$; Fig. \ref{expdpp} shows the delay scans for $\Delta (P1-P2)$= 20~ps, 200~ps , 1~ns, and 2~ns. The experimental conditions were chosen in such way that the I-phase forms as described previously: the temperature of the cryostat and the absorbed fluence related to P1 are set to 265 K and 4.6~mJ~cm$^{-2}$, respectively. The absorbed fluence for P2 used to excite the femtosecond dynamics of the forming I-CDW state was set to 2.3~mJ~cm$^{-2}$. The intensity oscillates with oscillation corresponding to the amplitude mode of the I-CDW. The intensity does not relax completely and therefore we use an adapted displacive excitation model for $t>t_0$ \cite{Zeiger1992,Laulhe2015}:
\begin{equation}
\frac{I(t)}{I_0} = (1+A_{d}[cos(2\pi \nu_{AM} t)e^{-\frac{t}{\tau_{AM}}}-d-(1-d)e^{-\frac{t}{\tau_{d}}}])^2
\end{equation}
where $A_d$, $\nu_{AM}$, and $\tau_{AM}$ represent  respectively the amplitude, the frequency,  and the damping time of the amplitude mode oscillations. The residual displacement $d$ and the decay time $\tau_{d}$ characterize the relaxation of the transient quasi-equilibrium atomic positions. The zero values for time were also fitted but did not change significantly because the setup only changed the P1 delay. The first drop of intensity is observed within 400 fs, a delay significantly larger than our time resolution (150 fs, determined using diffraction from Bi(111)\cite{Beaud2007}). The different fitted parameters are represented in Fig. \ref{expdpp} as a function of $\Delta (P1-P2)$. The damping time of the oscillation and its frequency do not change significantly with $\Delta (P1-P2)$. The amplitude of the oscillation and the loss of modulation amplitude at  4~ps,  however, both decrease with $\Delta (P1-P2)$, as does the recovery time $\tau_d$.  

Using the domain widths extracted from Fig. \ref{265K_40mW} and the fitted power law, we calculate the volume of the domains for each value of $\Delta (P1-P2)$. Fig. \ref{widthdpp} shows the amplitudes of the oscillation versus the domain volumes, which is fitted using a linear behavior, $A_d=aV +b$. We conclude that the magnitude of the coherently excited amplitude mode decreases as the I-CDW domains get larger.

\section{Discussion}

In our experiment for both the low and the high fluence, the increase of the intensity is entirely due to the volume increase of the domains, which suggests that there is no coalescence between I-phase regions. This constitutes a major difference with the results reported in Ref. \citep{Laulhe}, which show that at high fluences the photoinduced I phase regions coalesce, leading in turn to a coarsening behavior. The most significant difference between the two experiments is the photon energy of the pump, which is two times larger for the present experiment. 
The time-resolved electron diffraction work of Han \textit{et al.}~\cite{Han2015} showed very different threshold behavior for photoinduced phase transitions starting from the C-CDW phase for 800 nm versus 2500 nm pumping wavelengths, even after correcting for differences in absorption. This suggests that the occurrence of the photoinduced phase transitions does not depend on the overall absorbed energy density. In their article, Han \textit{et al.} rather suggest  a dependence on the number density of absorbed photons. The number of absorbed photons per unit cell can be estimated as:
\begin{equation}
n_p=\frac{F \Omega}{\delta_L \Delta E}
\end{equation}
where $F$ is the absorbed fluence, $\Omega$ is the unit cell volume, $\delta_L$ is the laser penetration depth, and $\Delta E$ is the photon energy. Table \ref{abph} shows the estimated density of absorbed photons for our experiment and for Refs. \cite{Laulhe,Vogelgesang2017}. Even though the photon energy in the experiment from Ref. \cite{Laulhe} is two times lower, the absorbed photons per unit cell is larger in our experiment. This would suggest that the lack of a coalescence time is not simply due to insufficient excitation levels, but is instead a consequence of the different electronic states excited by the two different wavelengths. Similar observations have been made previously \cite{Lantz2017,Ritschel2015,Yan-Bin2007}. In Ref.  \cite{Vogelgesang2017}, the photon energy was 1 eV but the absorbed photon density was much lower. Therefore we suggest that in order to observe a coarsening phenomenon, the photon energy has to be close to 0.8 eV and the absorbed photon density has to be high.

\begin{table}
\begin{tabular}{|l|c|c|c|}
  \hline
  ~& Laulh\'{e}  \textit{et al.}\cite{Laulhe} &Vogelgesang \textit{et al.} \cite{Vogelgesang2017}& this work\\
  \hline
 $\delta_L$ & 44 nm & 40 nm & 30 nm \\
 $\Delta E$ & 0.8 eV & 1 eV & 1.55 eV \\  
 $F$ & 6.9 mJ~cm$^{-2}$& 5.65 mJ~cm$^{-2}$& 10.5 mJ~cm$^{-2}$\\ 
 \hline
 $n_p$ & 0.71 & 0.51  & 0.81 \\  
  \hline
\end{tabular}
\caption{Number of photon absorbed per unit cell and data used to calculate it.}
\label{abph} 
\end{table}

The shapes of the domains are anisotropic. They are more elongated in the ab plane and shorter along c, (Figs. \ref{230K_90mW}\textbf{d} and \ref{265K_40mW}\textbf{d}), which is in accordance with the layered nature of the material. The difference in $a^*$ and $b^*$ directions are likely due to the mosaicity of the sample since the NC-satellite peak also has the width along $b^*$ larger than the one along $a^*$ (see Appendix A). The the I satellite peak widths exhibit the same relative decrease in all reciprocal space directions, which proves that for these fluences and this photon energy used the domain shape remains the same through out the growth process. Haupt \textit{et al.} suggested that the nucleation of the I-phase takes place in the discommensuration regions of the NC-phase. Since these regions are along $a^*$and $b^*$, the average observed distribution would also give dynamics that is the same along all directions. We therefore cannot on the basis of our data distinguish where the nucleation takes place. Because the growth process does not change the shape of the domains but only their sizes, it seems that the layered nature of 1T-TaS$_2$ does not limit the growth in the $c^*$ direction.  

The absolute values of the exponents of the power laws are higher for the high fluence than for the low fluence but still lower than the ones observed in Refs. \cite{Laulhe,Vogelgesang2017}. Fig. \ref{speed} shows the calculated growth velocities for the different directions and both fluences. We notice that the velocities are higher for the higher fluence.  In Ref. \cite{Laulhe}, at 100~ps the speed is 80~m~s$^{-1}$, which is higher than our observation. This velocity of domain growth is the same order of magnitude as the reorientation of ferromagnetic domains in magnetic materials \cite{Gorchon2014}. 

At the lower excitation fluence, the intensity drop after 1~ns is likely the result of the diffusion of heat and excited carriers into the bulk of the sample. As soon as the heat dissipates the temperature drops and the photoinduced I-phase disappears. However, the correlation length stays larger than during the growth. We suggest that the large domains and the domains with defects are more stable and therefore the intensity drop is mostly due to the disappearance of the small domains, which are less stable\cite{Abreu2015,Lantz2015}.

Using the double pump scheme, we observed that the dynamics of the CDW in the I-phase is sensitive to the volume size of its domains. The magnitude of the coherently excited I-CDW amplitude mode decreases with increasing volume of the CDW domains. Similarly, Mozurkewich and Gr\"{u}ner showed that the oscillation amplitude of the current density, which relates to the phase mode of the CDW, decreases with increasing volume \cite{Mozurkewich1983,Ishiguro1991}. Based on these observations, we hypothesize that in general the excitation modes of the CDW in response to a given perturbation increase in amplitude as the domain size decreases. The frequency of the amplitude mode is almost constant for all domain sizes, which shows that the fundamental eigen-modes are not affected by the size of the domains. The correlation length along the c$^*$ direction is around 10~nm, which is lower than the attenuation depth of the pump pulse (30~nm) as well as the attenuation length of the x-ray (130~nm), therefore the diffracted intensity arises from multiple pumped domains and we can exclude an artifact arising from an averaging mismatch.

\section{Conclusion}

We have shown that the photoinduced I-phase domains have different sizes depending on the lattice directions, and grow in a self-similar manner. Compared to previous experiments, no signature of the coarsening phenomenon was observed. This may be a consequence of using a different photon energy to drive the phase transition, although further more systematic investigations are needed to verify this hypothesis.

We also observed, using a double pump pulse scheme, a dependence of the newly created I-phase dynamics on the size of the domains. The amplitude of the oscillations is linearly related to the volume of the CDW domains, whereas their frequency is not affected. This suggests that larger domains are more stable with respect to electronic perturbations.

\appendix
    \section{Appendix A: Instrumental resolution}
The instrumental resolution is crucial for a correct observation of the widths of the diffracted peaks. The energy resolution gives a theoretical momentum resolution of 3.5~10$^{-4}$ \AA$^{-1}$, which is 7.5~10$^{-4}$ in reciprocal lattice units (r.l.u.) along h and k, and 3.7~10$^{-4}$ r.l.u along l. The geometry and detector pixel size can deteriorate the resolution even further. The detector pixel size is 172~$\mu$m and is placed at 9.7~cm from the sample. The peak observed was 5$^\circ$ in $\delta$ and 45.7$^\circ$ in $\gamma$, using the angle convention from Ref. \cite{Willmott2013}. Furthermore the measurement geometry was performed in a grazing incidence with a beam size 5$\times$200 $\mu$m$^2$, which leads to a footprint of the x-rays on the sample of 286$\times$200 $\mu$m$^2$. Taking all these different effects into consideration, we estimate that the resolution is 1.0 10$^{-3}$ r.l.u along h and k, and 1.1 10$^{-3}$ along l. To confirm this resolution we performed a reciprocal space mapping of the NC-satellite peak at equilibrium, which has exactly the same outgoing angles than the I-satellite peak. We find widths of $1.2\times1.6\times10.2$ 10$^{-3}$ in h$\times$k$\times$l r.l.u , which proves that we can resolve narrower widths than the ones of the I-satellite peak. However we consider the widths of the NC peak as an upper bound on the momentum resolution since the sample has a non-zero mosaicity which will increase the widths especially along the l direction.

{\em Acknowledgments.} We wish to warmly thank Sabrina Salmon for her valuable help during sample synthesis. Time-resolved x-ray diffraction measurements were carried out at the X05LA beam line of the Swiss Light Source, Paul Scherrer Institut, Villigen. We acknowledge financial support by the NCCR Molecular Ultrafast Science and Technology (NCCR MUST), a research instrument of the Swiss National Science Foundation (SNSF). G. L. also acknowledges the financial support of ETH Career Seed Grant SEED-80 16-1. E. A. acknowledges support from the ETH Zurich Postdoctoral Fellowship Program and from the Marie Curie Actions for People COFUND Program.

\bibliographystyle{apsrev}

\end{document}